\documentclass[reprint,superscriptaddress,
prb,aps,
]{revtex4-2}
\setcitestyle{super} 
\usepackage{graphicx}
\usepackage{dcolumn}
\usepackage{bm}
\usepackage[version=4]{mhchem}
\usepackage{ragged2e}
\usepackage{booktabs}

\usepackage{siunitx}
\DeclareSIUnit\bar{bar}
\begin{document}
\title{The graphene squeeze-film microphone}

\author{M.P.~Abrahams}
\affiliation{Department of Precision and Microsystems Engineering, Delft University of Technology, Mekelweg 2, 2628 CD Delft, The Netherlands, EU}

\author{J.~Martinez}
\affiliation{Multimedia Computing Group, Intelligent Systems Department, Faculty of Electrical Engineering, Mathematics and Computer Science, Delft University of Technology, 2628 XE Delft, The Netherlands, EU}

\author{P.G.~Steeneken}
\affiliation{Department of Precision and Microsystems Engineering, Delft University of Technology, Mekelweg 2, 2628 CD Delft, The Netherlands, EU}

\author{G.J.~Verbiest}
\affiliation{Department of Precision and Microsystems Engineering, Delft University of Technology, Mekelweg 2, 2628 CD Delft, The Netherlands, EU}
\email{G.J.Verbiest@tudelft.nl}



\begin{abstract}
Most microphones operate by detecting the sound-pressure induced motion of a membrane. In contrast, here we introduce a microphone that operates by monitoring the sound-pressure-induced modulation of the compressibility of air. By driving a graphene membrane at its resonance frequency, the gas, that is trapped in a squeeze-film beneath it, is compressed at high frequency. Since the stiffness of the gas film depend on the air pressure, the resonance frequency of the graphene is modulated by variations in sound pressure. We demonstrate that this squeeze-film microphone principle can be used to detect sound and music by tracking the membrane's resonance frequency using a phase-locked loop (PLL). Since the sound detection principle is different from conventional devices, the squeeze-film microphone potentially offers advantages like increased dynamic range, and a lower susceptibility to pressure-induced failure and vibration-induced noise. Moreover, it might be made much smaller, as demonstrated by the microphone in this work that operates using a circular graphene membrane with an area that is more than a factor 1000 smaller than that of MEMS microphones.
\end{abstract}

\maketitle

\begin{figure*}[t]
\begin{center}
\includegraphics[draft=false,keepaspectratio=true,clip,width=1.0\linewidth]{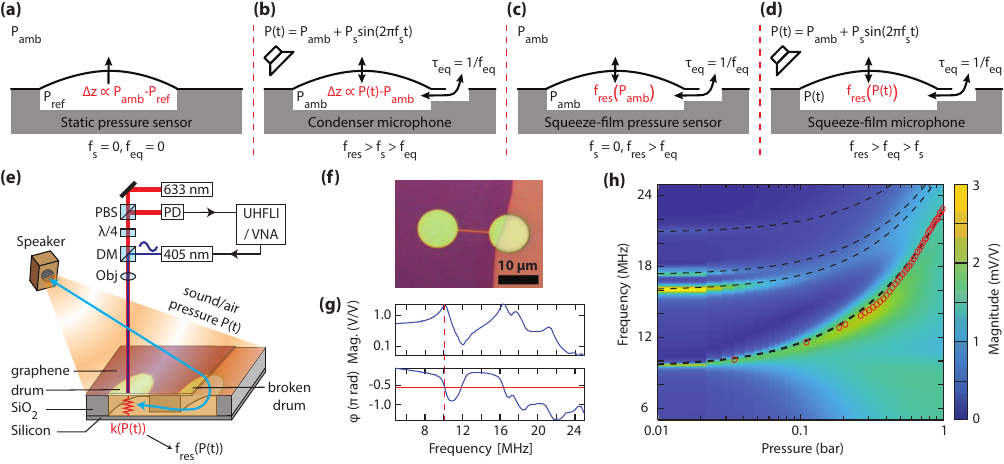}
\end{center}
\caption{Schematic drawings of (a) static pressure sensor, (b) condenser microphone, (c) squeeze-film pressure sensor, and (d) squeeze-film microphone. The squeeze-film devices distinguish them from conventional devices by having a pressure below the membrane that is always equal to the external pressure, caused by the relatively short equilibration time $\tau_\textrm{eq}$. Instead of measuring pressure by monitoring the deflection $\Delta z$ of the membrane, the squeeze-film devices determine gas pressure by monitoring its effect on the membrane's resonance frequency $f_{\rm res}$. (e) Illustration of a graphene drum over a cavity in a SiO$_2$/Si substrate. A red (633~nm) laser beam passes a polarized beam splitter (PBS) and a quarter-wave plate ($\lambda/4$) such that the reflected beam is diverted into a photodetector (PD) for analysis by a phase-locked loop (PLL, implemented in UHFLI) or vector network analyzer (VNA). A blue (405~nm) laser beam enters the red laser beam path through a dichroic mirror (DM) and excites the drum at a given frequency. The objective (Obj) focuses both laser beams on the drum. A speaker modulates the air pressure $P(t)$ by sound waves, which in turn modulate the spring constant $k$ of the drum and thus its resonance frequency $f_{\rm res}$. (f) An optical image of a typical device shows a suspended drum that is connected to the environment by a venting channel. (g) The magnitude and phase as a function of frequency measured by the VNA at a pressure of 10 mbar. The continuous red line indicates the $-\pi/2$ phase shift at the resonance frequency indicated by the dashed red line. (h) Magnitude as a function of frequency and pressure as recorded with the VNA. Red circles represent the extracted resonance frequencies and the black dashed lines fits to equation~\ref{eq1}.
\label{fig1}}
\end{figure*}

During the last centuries, a variety of membrane based devices have been developed to detect sound and pressure \cite{Malcovati2018,Zawawi2020,AliShah2018}. In static pressure sensors (Fig.~\ref{fig1}a) the membrane deflection is a measure of the difference between the outside pressure $P_{\rm amb}$ and the gas pressure $P_{\rm amb}$ in a sealed reference cavity. Condenser microphones (Fig.~\ref{fig1}b) operate by a similar principle, however they contain a small perforation, which causes static pressure differences to equilibriate at a time constant $\tau_{\rm eq}$, such that they only respond to pressure variations at sound frequencies $f_s>1/\tau_{\rm eq}$. 

A more recent concept is the squeeze-film pressure sensor\cite{Andrews1993,Smith2013} (Fig.~\ref{fig1}c), which consists of a membrane that is separated from a back-plate by a narrow gap with thickness $g_0$ that contains a thin film of gas which is at the same average pressure $P_{\rm amb}$ as the surrounding gas. The effective stiffness $k_{\rm eff}$ of the membrane does not only depend on its mechanical properties, but also on the compressibility of the gas. Since this compressibility is pressure dependent, the sensor's resonance frequency $f_{\rm res}=\frac{1}{2\pi} \sqrt{k_{\rm eff}/m_{\rm eff}}$, in which $m_{\rm eff}$ is the effective mass of the membrane, depends on pressure according to the following equation\cite{Dolleman2016}:

\begin{equation}%
f^2_{\rm res}(t)=f_0^2 + \frac{P(t)}{4\pi^2 g_{\textrm{0}}\rho h},%
\label{eq1}%
\end{equation}%
where $\rho$ is the membrane's mass density, $h$ its thickness, and $f_0$ its resonance frequency in vacuum.

Here we introduce and investigate a microphone that detects sound using the squeeze-film effect (Fig.~\ref{fig1}d), which we also patented \cite{Verbiest_patent_2020}. The operation principle of the squeeze-film microphone is based on the use of Eq.~(\ref{eq1}) to determine the time-dependent sound pressure $P_{\rm s}(t)$. This requires monitoring the resonance frequency $f_{\rm res}(t)$ at the sound frequency $f_{\rm s}$. As shown in Fig.~\ref{fig1}d, when sound modulates the gas pressure as $P(t)=P_{\rm amb}+P_{\rm s} \sin(2 \pi f_{\rm s} t)$, the time-dependence of the resonance frequency is found to be:
\begin{eqnarray}
    f_{\rm res}(P(t)) & \approx& f_{\rm res}(P_{\rm amb}) + \frac{{\rm d} f_{\rm res}}{{\rm d} P} P_{\rm s} \sin(2\pi f_{\rm s} t) \\
    &\approx & f_{\rm res}(P_{\rm amb}) + \frac{P_{\rm s} \sin(2\pi f_{\rm s} t)}{8\pi^2 g_{\textrm{0}}\rho h f_{\rm res}(P_{\rm amb})}  
\end{eqnarray}
For the squeeze film to detect sound pressure variations, the time $\tau_{\rm eq}=1/f_{\rm eq}$ it takes for the pressure in the squeeze-film to equilibrate with the ambient gas, needs to be much shorter than the period of the sound $\tau_{\rm s}=1/f_{\rm s}$. Also, for the squeeze-film gas to have a substantial effect on the resonance frequency, the period of the sound needs to be long compared to the typical response time $Q/f_{\rm res}$ of the resonator, which is less than 1 $\mu$s for the devices under investigation. Combining these conditions, we find that we need to design the squeeze-film microphone to satisfy $f_{\rm s} < f_{\rm eq} < f_{\rm res}/Q$. 

Graphene is the ideal material to satisfy these conditions, especially because of its low mass per unit area and high flexibility, which typically results in high resonance frequencies $f_{\rm res}\approx$ 10 MHz for membranes with a diameter of 10 $\mu$m and a $Q$-factor of $\approx 3$ in ambient conditions and $\approx 150$ in vacuum conditions \cite{Lemme2020,Steeneken2021}. The time it takes to equilibrate the pressure below the membrane via a channel is typically\cite{Roslon2020} in the range of $\tau_{\rm eq}=1-10~\mu s$, corresponding to $f_{\rm eq}=100-1000$ kHz, while the audible spectrum is in the range of $f_s=20$ Hz- 20 kHz. Moreover, Eq.~\ref{eq1} shows that the low mass per area $\rho h$ of graphene results in a high sensitivity of the resonance frequency to pressure variations, especially if the gap $g_{\rm 0}$ is small.

To realize a squeeze-film microphone (Fig.~\ref{fig1}f), we exfoliate few-layer graphene over a $g_{\rm 0} = 285$~nm deep dumbbell cavity with a radius of 5~$\mu$m in SiO$_{\rm 2}$ (see details in Supplementary Information~1). In total we measured six graphene squeeze-film microphones. We characterize the devices in a photo-thermal setup \cite{Davidovikj2016,Liu2023}, in which an intensity modulated blue laser ($\lambda=$405~nm) generates a thermal expansion force on the graphene drum near its resonance frequency and a red laser ($\lambda=$633~nm) measures its motion (Fig.~\ref{fig1}e). For each device, we measure the membrane's frequency response (Fig.~\ref{fig1}g) as a function of air pressure at room temperature to confirm the behavior predicted by equation~(\ref{eq1}). We extract the resonance frequency from experimental data at a given pressure, see dashed red line in Fig.~\ref{fig1}g and dashed lines in Fig.~\ref{fig1}h. The slope of the resonance frequency versus pressure at 1.0 bar (Fig.~\ref{fig1}h), defines the sensitivity $S_f=\frac{\textrm{d}f_{\rm res}}{\textrm{d} P}$ to sound pressure waves. By taking the pressure derivative of Eq.~\ref{eq1} this sensitivity is found to be $S_{\rm f}=(8 \pi^2 f_{\rm res} g_{\rm 0} \rho h)^{-1}$. From the experimental data we obtain a typical sensitivity of $S_{\rm f}=200$ Hz/Pa.  From this value we find that a squeeze-film microphone can capture a normal conversation, which corresponds to a root-mean-square (rms) sound pressure level of about 0.04 Pa (=65 dB$_\textrm{SPL}$), if it can detect resonance frequency modulations of $\sim$8 Hz.


\begin{figure*}[t]
\begin{center}
\includegraphics[draft=false,keepaspectratio=true,clip,width=1.0\linewidth]{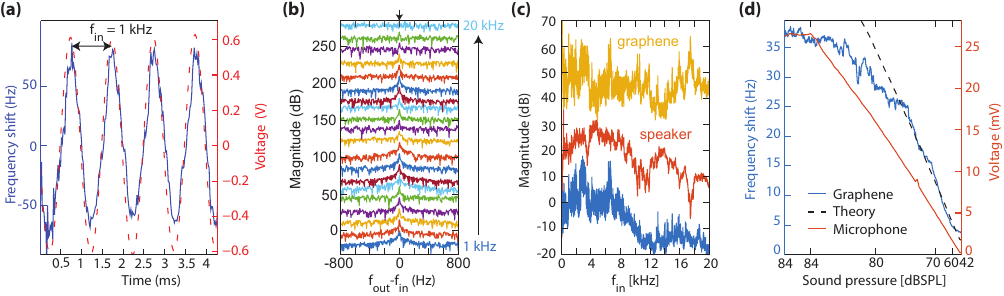}
\end{center}
\caption{(a) A time-snapshot of the shift in resonance frequency of the drum (blue) and the output of the reference microphone (red) due to a 1~kHz sound wave at 100~dB$_\textrm{SPL}$. (b) Waterfall plot of typical spectral responses of the drum around frequencies $f_{\rm out} = f_{\rm in}$ with $f_{\rm in}$ ranging from 1~kHz to 20~kHz. All curves are offsetted by 15~dB from one another. (c) The raw spectral response of the drum (blue) and of the speaker (red). After compensating for the speaker response, the graphene drum (yellow) shows a flat response for all audible frequencies. All curves are offsetted by 25~dB from one another. In panels (b) and (c), the dB-scale is obtained by using the amplitude of 68~Hz received at $f_{\rm in} = 1$~kHz and 100~dB$_\textrm{SPL}$ (see panel (a)) as reference. (d) Response of the graphene drum (blue) and reference microphone (red) to a 1~kHz sound wave for sound pressures ranging from 84~dB$_\textrm{SPL}$ to 42~dB$_\textrm{SPL}$. The black line indicates the response expected from equation~\ref{eq1}.
\label{fig2}}
\end{figure*}

After determining the pressure sensitivity of the resonance frequency, we use a Phase-Locked-Loop (PLL, Zurich Instruments UHFLI \cite{zhinst}) to track the resonance frequency of the graphene membrane as a function of time. We continuously update the frequency of the intensity modulation of the blue laser with the tracked resonance frequency to keep driving the graphene membrane at its resonance frequency.
Then, we generate sinusoidal sound waves at frequency $f_{\rm in}$ and sound pressure level $\sim$100~dB$_\textrm{SPL}=2.0$ Pa at the location of the microphone
with a speaker. We record the sound both with a reference microphone and with the graphene squeeze-film microphone. Fig.~\ref{fig2}a shows an exemplary measurement at $f_{\rm in} = 1$~kHz. The blue curve in the figure (left axis) shows the resonance frequency shift of the graphene microphone with respect to the nominal value of $\sim$22 MHz, due to the squeeze-film effect as determined by the PLL, while the red dashed line (right axis) shows the corresponding signal detected by the reference microphone. The correspondence between both signals provides evidence of the functionality of the squeeze-film microphone. We note that the response is somewhat lower than expected from the sensitivity of 200 Hz/Pa as estimated from Eq.~1.

We repeat the measurement at frequencies $f_{\rm in}$ ranging from 1~kHz
to 20~kHz and take the Fourier transform of the time signal coming from the PLL, like that in Fig.~\ref{fig2}a, to obtain the spectral responses depicted in Fig.~\ref{fig2}b. The spectra show a clear response of the graphene squeeze-film microphone at the input frequency $f_{\rm out} = f_{\rm in}$. The observed reduction of its output at 20 kHz can mainly be attributed to a reduction of the output power of the speaker at high frequencies (see Fig.~\ref{fig2}c). To make sure the detected sound signal is transmitted via air, and not generated by vibrations of the support of the microphone, we mounted the graphene devices on a vibrating piezo-electric transducer \cite{Verbiest2018}. By actuating the transducer at the membrane resonance frequency, we were not able to detect the motion and resonance of the membrane in air, providing evidence that the detected sound signal does not propagate through the sample support or substrate, but is transmitted via air (Supplementary Fig.~\ref{figS1}). By comparing the output of the graphene squeeze-film microphone to that of a calibrated reference microphone, we can obtain its transfer function, that is shown in Fig.~\ref{fig2}c. How we determined the transfer functions of the reference microphone and speaker is provided in Supplementary Information~2. As Fig.~\ref{fig2}c indicates and expected based on equation~\ref{eq1}, the sensitivity of the graphene membrane is roughly constant over the probed acoustic frequency range.

To determine the minimally detectable sound pressure level, we generate sinusoidal sound waves at frequency $f_{\rm in} = 1$~kHz and a strength varying from 84~$dB_{\rm SPL}$ to 42~$dB_{\rm SPL}$. Fig.~\ref{fig2}d shows that the response of the graphene device agrees nicely with the one expected based on equation~\ref{eq1}. Moreover, it follows the signal measured by the reference microphone quite well. Graphene devices can thus sense sound pressure levels down to 42~$dB_{\rm SPL}$.

\begin{figure}[t]
\begin{center}
\includegraphics[draft=false,keepaspectratio=true,clip,width=1.0\linewidth]{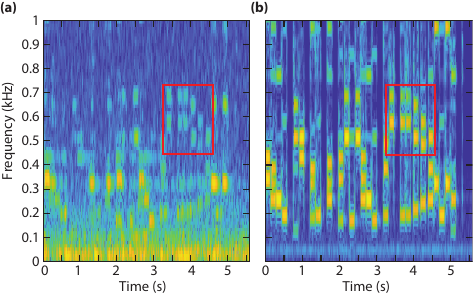}
\end{center}
\caption{Selected normalized spectrograms of the super mario theme song recorded with (a) the graphene squeeze-film microphone and (b) the reference microphone. The red squares enable direct comparison between data in panels (a) and (b).
\label{fig3}}
\end{figure}

The presented graphene devices even allow for the recording of music. A 5.5 seconds long spectrogram of the `Super Mario Theme Song' \cite{Kondo1985} recorded with a graphene device and with the reference microphone is shown in Fig.~\ref{fig3}a and \ref{fig3}b, respectively. Full recordings are provided as Supplementary Material. By comparing both panels of Fig.~\ref{fig3}, we observe the response of the graphene device can captures the main features of the sound signal below $\sim$750~Hz, although it contains more noise than the reference microphone. This shows the functionality of graphene membranes with a diameter of only 10~$\mu$m as microphone.

\begin{figure*}[hbt]
\begin{center}
\includegraphics[width=0.8\linewidth]{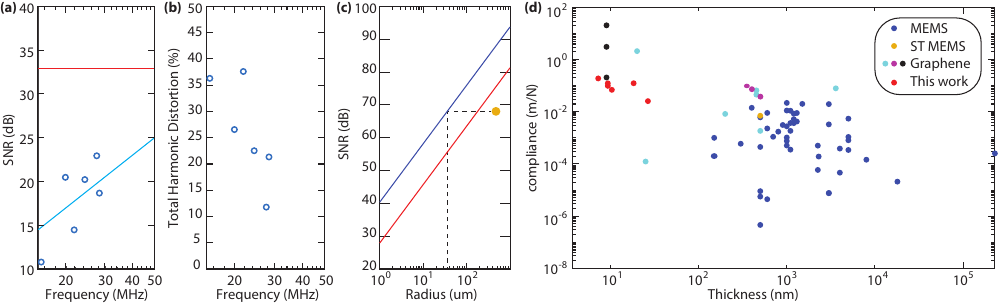}
\end{center}
\caption{(a) Signal-to-noise ratio and (b) total harmonic distortion of the measured squeeze-film microphones (cyan circles) as a function of their resonance frequency. The SNR increases approximately linearly with the logarithm of the resonance frequency (cyan line in a). The red line in a indicates the SNR limit by thermo-mechanical noise. (c) Thermo-mechanical SNR limit at a sound pressure level of 80~dB (red) and 94~dB (blue). The orange point marks the performance of the MP23DB01HP of STMicroelectronics. The black dashed lines indicate the possible miniaturization by squeeze-film microphones at constant SNR. (d) Comparison of the mechanical compliance and thickness of the presented squeeze-film microphones (red) to other MEMS microphones (blue) \cite{Zawawi2020}, to the MP23DB01HP of STMicroelectronics (orange), and different types of graphene microphones (cyan \cite{Siskins2020,Fan2019,Zhou2015,Wittmann2019,Wood2019}, magenta \cite{Todorovic2015,Woo2017,Li2015}, black \cite{Baglioni2023}).
\label{fig4}}
\end{figure*}

To benchmark the fabricated graphene devices, we determine the signal-to-noise ratio (SNR) at 1~kHz according to the procedure that is used for commercial microphones. This process is explained in detail in Supplementary information~3. In short, a continuous sound wave of 1~kHz at 80~dB$_{\rm SPL}$ excites the microphone. For 5 different devices, the graphene microphone response (resonance frequency shift as function of time) is recorded via the PLL and converted to the frequency domain using a Fourier transform. The signal strength is determined over a small bandwidth around 1~kHz, while the noise floor is also detected, like in Fig.~\ref{fig2}b. We also extract the strength of all harmonics of 1~kHz up to 20~kHz to determine the total harmonic distortion (THD). The rest of the spectrum is considered to be noise. The extracted SNR and THD values are listed in figures~\ref{fig4}a and \ref{fig4}b for all measured devices as a function of their resonance frequency $f_{\rm res}$ at atmospheric pressure, which varied due to tension and membrane thickness variations. As expected (Supplementary Information 4, Eq.~\ref{seq7}), the SNR increases proportional to the resonance frequency $f_{\rm 0}$ up to a maximum of 23 dB, which corresponds to a detection limit of $80-23=57$ dB$_\textrm{SPL}$. With increasing $f_{\rm res}$, also the THD decreases to a minimum of 12\%.  

The ultimate signal-to-noise ratio of the microphone is set by the thermo-mechanical motion of the graphene membrane. The squeeze-film effect translates this thermo-mechanical motion into frequency noise. As the signal is determined by equation~\ref{eq1} and the strength of the incoming pressure wave, the signal-to-noise ratio can be analytically computed (Supplementary information~4) and depends on the radius of the membrane, temperature, and ambient pressure. The outcome, for an incident sound pressure level of 80 dB$_\textrm{SPL}$, is indicated by the red line in Fig.~\ref{fig4}a. The squeeze-film microphones presented in this work are roughly 20~dB below the thermo-mechanical SNR limit (blue circles in Fig.~\ref{fig4}a), which might be attributed to the contribution of other sources and frequency fluctuations than thermo-mechanical (Brownian) noise. To show the potential of squeeze-film microphones, we show the calculated (Eq.~\ref{seq8}) thermo-mechanically limited SNR in Fig.~\ref{fig4}c for a sound pressure level of 80~dB (red) and 94~dB (blue) together with the specified SNR of the MP23DB01HP of STMicroelectronics \cite{st} for 94~dB$_\textrm{SPL}$ at 1~kHz. The MP23DB01HP uses a membrane radius of 475~$\mu$m to reach this SNR level. However, a squeeze-film microphone reaches a (thermomechanical noise limited) SNR value of 64~dB at ambient pressure and room temperature at an almost 15x smaller radius of 18~$\mu$m. The experimental total harmonic distortion (see SI 3) of the devices under study is shown in Fig.~\ref{fig4}b.

The sensitivity of conventional MEMS microphones is determined by the compliance and thickness of the membrane. A high compliance and small thickness is desired for best microphone performance. For the squeeze-film microphone, this does not hold. In the thermo-mechanical limit, both the signal and noise scale identically with the membrane thickness and compliance. Consequently, the achievable SNR does not directly depend on the compliance nor on the thickness. To reach the thermo-mechanical limit, we just require a high resonance frequency to maintain the condition $f_{\rm res}>f_{\rm eq}>f_{\rm s}$. This could be achieved by increasing the membrane tension or decreasing the membrane mass per area (e.g. by reducing the membrane thickness). This lifts design and fabrication constraints, as high tension membranes can be routinely fabricated in Si$_{\rm 3}$N$_{\rm 4}$ technology. Nevertheless, a high compliance can still be helpful, since although it does not affect the absolute sensitivity, it does improve the fractional sensitivity. In that respect the high compliance values obtained with the graphene membranes (see Fig.~\ref{fig4}d) are beneficial.

Compared to conventional microphones the squeeze-film pressure sensor has several advantages. First of all, theoretically its sensitivity is approximately frequency independent and only depends on the gap size $g_0$, the membrane size, tension, mass and dimensions according to Eq.~\ref{seq6}. This means that very low sound frequencies can be detected, and the microphone can even operate simultaneously as static pressure sensor by monitoring $f_{\rm res}$. The maximum sound frequency detectable by the squeeze film is limited by the frequency $f_{\rm eq}$ which can be of the order of 100-1000 kHz. The squeeze-film microphone might therefore also operate in the ultrasound regime, although it will become more challenging to track the resonance fast enough in this range. 
Further potential advantages of the squeeze-film microphone are its relative robustness against sudden external pressure changes, due to its short pressure equilibration time $\tau_{\rm eq}$ and its intrinsic flatband response, independent of frequency, based on Eq.~\ref{eq1}. Moreover, we expect the squeeze-film microphone to be extremely robust at very high sound pressure levels, because the sound pressure does not directly result in a force on the membrane like in conventional microphones. On the other hand, a main challenge in the operation of the squeeze-film microphone is its relatively complex readout methodology. It will require advanced circuit design developments to be able to realize a readout circuit that can compete with the low-power, low-cost CMOS circuits that are used to readout current MEMS microphones \cite{Wang2015}.

In summary, we present a new microphone concept utilizing the squeeze-film effect for detecting sound. We show that the microphone's resonance frequency is indeed quite sensitive to pressure changes, and that the sound-induced frequency changes due to the tension modulation can be detected by a PLL circuit. The low mass and high-flexibility of graphene make it very suitable for this type of pressure sensing. Although the readout of the microphone is more complex, it offers several advantages like broad bandwidth, small membrane size and potential robustness to high sound-levels, external vibrations, and sudden pressure changes. Ultimately, the squeeze-film microphone enables further down-scaling of microphone technology by at least an order of magnitude thus providing a sound sensing technology that, if brought to higher maturity level, has the potential to complement or partly replace current microphones.


\bibliography{main}

\begin{thebibliography}{31}%
\makeatletter
\providecommand \@ifxundefined [1]{%
 \@ifx{#1\undefined}
}%
\providecommand \@ifnum [1]{%
 \ifnum #1\expandafter \@firstoftwo
 \else \expandafter \@secondoftwo
 \fi
}%
\providecommand \@ifx [1]{%
 \ifx #1\expandafter \@firstoftwo
 \else \expandafter \@secondoftwo
 \fi
}%
\providecommand \natexlab [1]{#1}%
\providecommand \enquote  [1]{``#1''}%
\providecommand \bibnamefont  [1]{#1}%
\providecommand \bibfnamefont [1]{#1}%
\providecommand \citenamefont [1]{#1}%
\providecommand \href@noop [0]{\@secondoftwo}%
\providecommand \href [0]{\begingroup \@sanitize@url \@href}%
\providecommand \@href[1]{\@@startlink{#1}\@@href}%
\providecommand \@@href[1]{\endgroup#1\@@endlink}%
\providecommand \@sanitize@url [0]{\catcode `\\12\catcode `\$12\catcode
  `\&12\catcode `\#12\catcode `\^12\catcode `\_12\catcode `\%12\relax}%
\providecommand \@@startlink[1]{}%
\providecommand \@@endlink[0]{}%
\providecommand \url  [0]{\begingroup\@sanitize@url \@url }%
\providecommand \@url [1]{\endgroup\@href {#1}{\urlprefix }}%
\providecommand \urlprefix  [0]{URL }%
\providecommand \Eprint [0]{\href }%
\providecommand \doibase [0]{https://doi.org/}%
\providecommand \selectlanguage [0]{\@gobble}%
\providecommand \bibinfo  [0]{\@secondoftwo}%
\providecommand \bibfield  [0]{\@secondoftwo}%
\providecommand \translation [1]{[#1]}%
\providecommand \BibitemOpen [0]{}%
\providecommand \bibitemStop [0]{}%
\providecommand \bibitemNoStop [0]{.\EOS\space}%
\providecommand \EOS [0]{\spacefactor3000\relax}%
\providecommand \BibitemShut  [1]{\csname bibitem#1\endcsname}%
\let\auto@bib@innerbib\@empty
\bibitem [{\citenamefont {Malcovati}\ and\ \citenamefont
  {Baschirotto}(2018)}]{Malcovati2018}%
  \BibitemOpen
  \bibfield  {author} {\bibinfo {author} {\bibfnamefont {P.}~\bibnamefont
  {Malcovati}}\ and\ \bibinfo {author} {\bibfnamefont {A.}~\bibnamefont
  {Baschirotto}},\ }\bibfield  {journal} {\bibinfo  {journal} {Micromachines}\
  }\textbf {\bibinfo {volume} {9}},\ \href {https://doi.org/10.3390/mi9070323}
  {10.3390/mi9070323} (\bibinfo {year} {2018})\BibitemShut {NoStop}%
\bibitem [{\citenamefont {Zawawi}\ \emph {et~al.}(2020)\citenamefont {Zawawi},
  \citenamefont {Hamzah}, \citenamefont {Majlis},\ and\ \citenamefont
  {Mohd-Yasin}}]{Zawawi2020}%
  \BibitemOpen
  \bibfield  {author} {\bibinfo {author} {\bibfnamefont {S.~A.}\ \bibnamefont
  {Zawawi}}, \bibinfo {author} {\bibfnamefont {A.~A.}\ \bibnamefont {Hamzah}},
  \bibinfo {author} {\bibfnamefont {B.~Y.}\ \bibnamefont {Majlis}},\ and\
  \bibinfo {author} {\bibfnamefont {F.}~\bibnamefont {Mohd-Yasin}},\ }\bibfield
   {journal} {\bibinfo  {journal} {Micromachines}\ }\textbf {\bibinfo {volume}
  {11}},\ \href {https://doi.org/10.3390/mi11050484} {10.3390/mi11050484}
  (\bibinfo {year} {2020})\BibitemShut {NoStop}%
\bibitem [{\citenamefont {Ali~Shah}\ \emph {et~al.}(2018)\citenamefont
  {Ali~Shah}, \citenamefont {Ali~Shah}, \citenamefont {Lee},\ and\
  \citenamefont {Hur}}]{AliShah2018}%
  \BibitemOpen
  \bibfield  {author} {\bibinfo {author} {\bibfnamefont {M.}~\bibnamefont
  {Ali~Shah}}, \bibinfo {author} {\bibfnamefont {I.}~\bibnamefont {Ali~Shah}},
  \bibinfo {author} {\bibfnamefont {D.-G.}\ \bibnamefont {Lee}},\ and\ \bibinfo
  {author} {\bibfnamefont {S.}~\bibnamefont {Hur}},\ }\bibfield  {journal}
  {\bibinfo  {journal} {Journal of Sensors}\ }\href
  {https://doi.org/10.1155/2019/9294528} {10.1155/2019/9294528} (\bibinfo
  {year} {2018})\BibitemShut {NoStop}%
\bibitem [{\citenamefont {Andrews}\ \emph {et~al.}(1993)\citenamefont
  {Andrews}, \citenamefont {Turner}, \citenamefont {Harris},\ and\
  \citenamefont {Harris}}]{Andrews1993}%
  \BibitemOpen
  \bibfield  {author} {\bibinfo {author} {\bibfnamefont {M.}~\bibnamefont
  {Andrews}}, \bibinfo {author} {\bibfnamefont {G.}~\bibnamefont {Turner}},
  \bibinfo {author} {\bibfnamefont {P.}~\bibnamefont {Harris}},\ and\ \bibinfo
  {author} {\bibfnamefont {I.}~\bibnamefont {Harris}},\ }\href
  {https://doi.org/https://doi.org/10.1016/0924-4247(93)80196-N} {\bibfield
  {journal} {\bibinfo  {journal} {Sensors and Actuators A: Physical}\ }\textbf
  {\bibinfo {volume} {36}},\ \bibinfo {pages} {219} (\bibinfo {year}
  {1993})}\BibitemShut {NoStop}%
\bibitem [{\citenamefont {Smith}\ \emph {et~al.}(2013)\citenamefont {Smith},
  \citenamefont {Niklaus}, \citenamefont {Paussa}, \citenamefont {Vaziri},
  \citenamefont {Fischer}, \citenamefont {Sterner}, \citenamefont {Forsberg},
  \citenamefont {Delin}, \citenamefont {Esseni}, \citenamefont {Palestri},
  \citenamefont {Östling},\ and\ \citenamefont {Lemme}}]{Smith2013}%
  \BibitemOpen
  \bibfield  {author} {\bibinfo {author} {\bibfnamefont {A.~D.}\ \bibnamefont
  {Smith}}, \bibinfo {author} {\bibfnamefont {F.}~\bibnamefont {Niklaus}},
  \bibinfo {author} {\bibfnamefont {A.}~\bibnamefont {Paussa}}, \bibinfo
  {author} {\bibfnamefont {S.}~\bibnamefont {Vaziri}}, \bibinfo {author}
  {\bibfnamefont {A.~C.}\ \bibnamefont {Fischer}}, \bibinfo {author}
  {\bibfnamefont {M.}~\bibnamefont {Sterner}}, \bibinfo {author} {\bibfnamefont
  {F.}~\bibnamefont {Forsberg}}, \bibinfo {author} {\bibfnamefont
  {A.}~\bibnamefont {Delin}}, \bibinfo {author} {\bibfnamefont
  {D.}~\bibnamefont {Esseni}}, \bibinfo {author} {\bibfnamefont
  {P.}~\bibnamefont {Palestri}}, \bibinfo {author} {\bibfnamefont
  {M.}~\bibnamefont {Östling}},\ and\ \bibinfo {author} {\bibfnamefont
  {M.~C.}\ \bibnamefont {Lemme}},\ }\href {https://doi.org/10.1021/nl401352k}
  {\bibfield  {journal} {\bibinfo  {journal} {Nano Letters}\ }\textbf {\bibinfo
  {volume} {13}},\ \bibinfo {pages} {3237} (\bibinfo {year} {2013})},\ \bibinfo
  {note} {pMID: 23786215},\ \Eprint
  {https://arxiv.org/abs/https://doi.org/10.1021/nl401352k}
  {https://doi.org/10.1021/nl401352k} \BibitemShut {NoStop}%
\bibitem [{\citenamefont {Dolleman}\ \emph {et~al.}(2016)\citenamefont
  {Dolleman}, \citenamefont {Davidovikj}, \citenamefont {Cartamil-Bueno},
  \citenamefont {van~der Zant},\ and\ \citenamefont
  {Steeneken}}]{Dolleman2016}%
  \BibitemOpen
  \bibfield  {author} {\bibinfo {author} {\bibfnamefont {R.~J.}\ \bibnamefont
  {Dolleman}}, \bibinfo {author} {\bibfnamefont {D.}~\bibnamefont
  {Davidovikj}}, \bibinfo {author} {\bibfnamefont {S.~J.}\ \bibnamefont
  {Cartamil-Bueno}}, \bibinfo {author} {\bibfnamefont {H.~S.~J.}\ \bibnamefont
  {van~der Zant}},\ and\ \bibinfo {author} {\bibfnamefont {P.~G.}\ \bibnamefont
  {Steeneken}},\ }\href {https://doi.org/10.1021/acs.nanolett.5b04251}
  {\bibfield  {journal} {\bibinfo  {journal} {Nano Letters}\ }\textbf {\bibinfo
  {volume} {16}},\ \bibinfo {pages} {568} (\bibinfo {year} {2016})},\ \bibinfo
  {note} {pMID: 26695136},\ \Eprint
  {https://arxiv.org/abs/https://doi.org/10.1021/acs.nanolett.5b04251}
  {https://doi.org/10.1021/acs.nanolett.5b04251} \BibitemShut {NoStop}%
\bibitem [{\citenamefont {Verbiest}\ \emph {et~al.}(2022)\citenamefont
  {Verbiest}, \citenamefont {Steeneken}, \citenamefont {Abrahams},\ and\
  \citenamefont {Martinez}}]{Verbiest_patent_2020}%
  \BibitemOpen
  \bibfield  {author} {\bibinfo {author} {\bibfnamefont {G.}~\bibnamefont
  {Verbiest}}, \bibinfo {author} {\bibfnamefont {P.}~\bibnamefont {Steeneken}},
  \bibinfo {author} {\bibfnamefont {M.}~\bibnamefont {Abrahams}},\ and\
  \bibinfo {author} {\bibfnamefont {C.}~\bibnamefont {Martinez}},\ }\href@noop
  {} {\bibfield  {journal} {\bibinfo  {journal} {patent NL2026284B1;
  WO2022039596A1}\ } (\bibinfo {year} {2022})}\BibitemShut {NoStop}%
\bibitem [{\citenamefont {Lemme}\ \emph {et~al.}(2020)\citenamefont {Lemme},
  \citenamefont {Wagner}, \citenamefont {Lee}, \citenamefont {Fan},
  \citenamefont {Verbiest}, \citenamefont {Wittmann}, \citenamefont {Lukas},
  \citenamefont {Dolleman}, \citenamefont {Niklaus}, \citenamefont {van~der
  Zant}, \citenamefont {Duesberg},\ and\ \citenamefont
  {Steeneken}}]{Lemme2020}%
  \BibitemOpen
  \bibfield  {author} {\bibinfo {author} {\bibfnamefont {M.~C.}\ \bibnamefont
  {Lemme}}, \bibinfo {author} {\bibfnamefont {S.}~\bibnamefont {Wagner}},
  \bibinfo {author} {\bibfnamefont {K.}~\bibnamefont {Lee}}, \bibinfo {author}
  {\bibfnamefont {X.}~\bibnamefont {Fan}}, \bibinfo {author} {\bibfnamefont
  {G.~J.}\ \bibnamefont {Verbiest}}, \bibinfo {author} {\bibfnamefont
  {S.}~\bibnamefont {Wittmann}}, \bibinfo {author} {\bibfnamefont
  {S.}~\bibnamefont {Lukas}}, \bibinfo {author} {\bibfnamefont {R.~J.}\
  \bibnamefont {Dolleman}}, \bibinfo {author} {\bibfnamefont {F.}~\bibnamefont
  {Niklaus}}, \bibinfo {author} {\bibfnamefont {H.~S.~J.}\ \bibnamefont
  {van~der Zant}}, \bibinfo {author} {\bibfnamefont {G.~S.}\ \bibnamefont
  {Duesberg}},\ and\ \bibinfo {author} {\bibfnamefont {P.~G.}\ \bibnamefont
  {Steeneken}},\ }\href {https://doi.org/10.34133/2020/8748602} {\bibfield
  {journal} {\bibinfo  {journal} {Research}\ }\textbf {\bibinfo {volume}
  {2020}} (\bibinfo {year} {2020})},\ \Eprint
  {https://arxiv.org/abs/https://spj.science.org/doi/pdf/10.34133/2020/8748602}
  {https://spj.science.org/doi/pdf/10.34133/2020/8748602} \BibitemShut
  {NoStop}%
\bibitem [{\citenamefont {Steeneken}\ \emph {et~al.}(2021)\citenamefont
  {Steeneken}, \citenamefont {Dolleman}, \citenamefont {Davidovikj},
  \citenamefont {Alijani},\ and\ \citenamefont {van~der Zant}}]{Steeneken2021}%
  \BibitemOpen
  \bibfield  {author} {\bibinfo {author} {\bibfnamefont {P.~G.}\ \bibnamefont
  {Steeneken}}, \bibinfo {author} {\bibfnamefont {R.~J.}\ \bibnamefont
  {Dolleman}}, \bibinfo {author} {\bibfnamefont {D.}~\bibnamefont
  {Davidovikj}}, \bibinfo {author} {\bibfnamefont {F.}~\bibnamefont
  {Alijani}},\ and\ \bibinfo {author} {\bibfnamefont {H.~S.~J.}\ \bibnamefont
  {van~der Zant}},\ }\href {https://doi.org/10.1088/2053-1583/ac152c}
  {\bibfield  {journal} {\bibinfo  {journal} {2D Materials}\ }\textbf {\bibinfo
  {volume} {8}},\ \bibinfo {pages} {042001} (\bibinfo {year}
  {2021})}\BibitemShut {NoStop}%
\bibitem [{\citenamefont {Roslon}\ \emph {et~al.}(2020)\citenamefont {Roslon},
  \citenamefont {Dolleman}, \citenamefont {Licona}, \citenamefont {Lee},
  \citenamefont {Siskins}, \citenamefont {Lebius}, \citenamefont {Madau\ss},
  \citenamefont {Schleberger}, \citenamefont {Alijani}, \citenamefont {van~der
  Zant},\ and\ \citenamefont {Steeneken}}]{Roslon2020}%
  \BibitemOpen
  \bibfield  {author} {\bibinfo {author} {\bibfnamefont {I.~E.}\ \bibnamefont
  {Roslon}}, \bibinfo {author} {\bibfnamefont {R.~J.}\ \bibnamefont
  {Dolleman}}, \bibinfo {author} {\bibfnamefont {H.}~\bibnamefont {Licona}},
  \bibinfo {author} {\bibfnamefont {M.}~\bibnamefont {Lee}}, \bibinfo {author}
  {\bibfnamefont {M.}~\bibnamefont {Siskins}}, \bibinfo {author} {\bibfnamefont
  {H.}~\bibnamefont {Lebius}}, \bibinfo {author} {\bibnamefont {Madau\ss}},
  \bibinfo {author} {\bibfnamefont {M.}~\bibnamefont {Schleberger}}, \bibinfo
  {author} {\bibfnamefont {F.}~\bibnamefont {Alijani}}, \bibinfo {author}
  {\bibfnamefont {H.~S.~J.}\ \bibnamefont {van~der Zant}},\ and\ \bibinfo
  {author} {\bibfnamefont {P.~G.}\ \bibnamefont {Steeneken}},\ }\href
  {https://doi.org/10.1038/s41467-020-19893-5} {\bibfield  {journal} {\bibinfo
  {journal} {Nature Communications}\ }\textbf {\bibinfo {volume} {11}},\
  \bibinfo {pages} {1} (\bibinfo {year} {2020})}\BibitemShut {NoStop}%
\bibitem [{\citenamefont {Davidovikj}\ \emph {et~al.}(2016)\citenamefont
  {Davidovikj}, \citenamefont {Slim}, \citenamefont {Cartamil-Bueno},
  \citenamefont {van~der Zant}, \citenamefont {Steeneken},\ and\ \citenamefont
  {Venstra}}]{Davidovikj2016}%
  \BibitemOpen
  \bibfield  {author} {\bibinfo {author} {\bibfnamefont {D.}~\bibnamefont
  {Davidovikj}}, \bibinfo {author} {\bibfnamefont {J.~J.}\ \bibnamefont
  {Slim}}, \bibinfo {author} {\bibfnamefont {S.~J.}\ \bibnamefont
  {Cartamil-Bueno}}, \bibinfo {author} {\bibfnamefont {H.~S.~J.}\ \bibnamefont
  {van~der Zant}}, \bibinfo {author} {\bibfnamefont {P.~G.}\ \bibnamefont
  {Steeneken}},\ and\ \bibinfo {author} {\bibfnamefont {W.~J.}\ \bibnamefont
  {Venstra}},\ }\href {https://doi.org/10.1021/acs.nanolett.6b00477} {\bibfield
   {journal} {\bibinfo  {journal} {Nano Letters}\ }\textbf {\bibinfo {volume}
  {16}},\ \bibinfo {pages} {2768} (\bibinfo {year} {2016})},\ \bibinfo {note}
  {pMID: 26954525},\ \Eprint
  {https://arxiv.org/abs/https://doi.org/10.1021/acs.nanolett.6b00477}
  {https://doi.org/10.1021/acs.nanolett.6b00477} \BibitemShut {NoStop}%
\bibitem [{\citenamefont {Liu}\ \emph {et~al.}(2023{\natexlab{a}})\citenamefont
  {Liu}, \citenamefont {Lee}, \citenamefont {\ifmmode \check{S}\else
  \v{S}\fi{}i\ifmmode~\check{s}\else \v{s}\fi{}kins}, \citenamefont {van~der
  Zant}, \citenamefont {Steeneken},\ and\ \citenamefont {Verbiest}}]{Liu2023}%
  \BibitemOpen
  \bibfield  {author} {\bibinfo {author} {\bibfnamefont {H.}~\bibnamefont
  {Liu}}, \bibinfo {author} {\bibfnamefont {M.}~\bibnamefont {Lee}}, \bibinfo
  {author} {\bibfnamefont {M.}~\bibnamefont {\ifmmode \check{S}\else
  \v{S}\fi{}i\ifmmode~\check{s}\else \v{s}\fi{}kins}}, \bibinfo {author}
  {\bibfnamefont {H.~S.~J.}\ \bibnamefont {van~der Zant}}, \bibinfo {author}
  {\bibfnamefont {P.~G.}\ \bibnamefont {Steeneken}},\ and\ \bibinfo {author}
  {\bibfnamefont {G.~J.}\ \bibnamefont {Verbiest}},\ }\href
  {https://doi.org/10.1103/PhysRevB.108.L081401} {\bibfield  {journal}
  {\bibinfo  {journal} {Phys. Rev. B}\ }\textbf {\bibinfo {volume} {108}},\
  \bibinfo {pages} {L081401} (\bibinfo {year}
  {2023}{\natexlab{a}})}\BibitemShut {NoStop}%
\bibitem [{\citenamefont {https://www.zhinst.com}()}]{zhinst}%
  \BibitemOpen
  \bibfield  {author} {\bibinfo {author} {\bibnamefont
  {https://www.zhinst.com}},\ }\href@noop {} {\bibinfo  {journal} {UHFLI 600MHz
  lock-in amplifier}\ }\BibitemShut {NoStop}%
\bibitem [{\citenamefont {Verbiest}\ \emph {et~al.}(2018)\citenamefont
  {Verbiest}, \citenamefont {Kirchhof}, \citenamefont {Sonntag}, \citenamefont
  {Goldsche}, \citenamefont {Khodkov},\ and\ \citenamefont
  {Stampfer}}]{Verbiest2018}%
  \BibitemOpen
\bibfield  {journal} {  }\bibfield  {author} {\bibinfo {author} {\bibfnamefont
  {G.~J.}\ \bibnamefont {Verbiest}}, \bibinfo {author} {\bibfnamefont {J.~N.}\
  \bibnamefont {Kirchhof}}, \bibinfo {author} {\bibfnamefont {J.}~\bibnamefont
  {Sonntag}}, \bibinfo {author} {\bibfnamefont {M.}~\bibnamefont {Goldsche}},
  \bibinfo {author} {\bibfnamefont {T.}~\bibnamefont {Khodkov}},\ and\ \bibinfo
  {author} {\bibfnamefont {C.}~\bibnamefont {Stampfer}},\ }\href
  {https://doi.org/10.1021/acs.nanolett.8b02036} {\bibfield  {journal}
  {\bibinfo  {journal} {Nano Letters}\ }\textbf {\bibinfo {volume} {18}},\
  \bibinfo {pages} {5132} (\bibinfo {year} {2018})},\ \bibinfo {note} {pMID:
  29989827},\ \Eprint
  {https://arxiv.org/abs/https://doi.org/10.1021/acs.nanolett.8b02036}
  {https://doi.org/10.1021/acs.nanolett.8b02036} \BibitemShut {NoStop}%
\bibitem [{\citenamefont {Kondo}(1985)}]{Kondo1985}%
  \BibitemOpen
  \bibfield  {author} {\bibinfo {author} {\bibfnamefont {K.}~\bibnamefont
  {Kondo}},\ }\href@noop {} {\bibfield  {journal} {\bibinfo  {journal}
  {Nintendo, HAL Laboratory, Inc.}\ } (\bibinfo {year} {1985})}\BibitemShut
  {NoStop}%
\bibitem [{\citenamefont {https://www.st.com/en/mems-and
  sensors/mp23db01hp}()}]{st}%
  \BibitemOpen
  \bibfield  {author} {\bibinfo {author} {\bibnamefont
  {https://www.st.com/en/mems-and sensors/mp23db01hp}},\ }\href@noop {}
  {}\BibitemShut {NoStop}%
\bibitem [{\citenamefont {Siskins}\ \emph {et~al.}(2020)\citenamefont
  {Siskins}, \citenamefont {Lee}, \citenamefont {Wehenkel}, \citenamefont {van
  Rijn}, \citenamefont {de~Jong}, \citenamefont {Renshof}, \citenamefont
  {Hopman}, \citenamefont {Peters}, \citenamefont {Davidovikj}, \citenamefont
  {van~der Zant},\ and\ \citenamefont {Steeneken}}]{Siskins2020}%
  \BibitemOpen
  \bibfield  {author} {\bibinfo {author} {\bibfnamefont {M.}~\bibnamefont
  {Siskins}}, \bibinfo {author} {\bibfnamefont {M.}~\bibnamefont {Lee}},
  \bibinfo {author} {\bibfnamefont {D.}~\bibnamefont {Wehenkel}}, \bibinfo
  {author} {\bibfnamefont {R.}~\bibnamefont {van Rijn}}, \bibinfo {author}
  {\bibfnamefont {T.}~\bibnamefont {de~Jong}}, \bibinfo {author} {\bibfnamefont
  {J.}~\bibnamefont {Renshof}}, \bibinfo {author} {\bibfnamefont
  {B.}~\bibnamefont {Hopman}}, \bibinfo {author} {\bibfnamefont
  {W.}~\bibnamefont {Peters}}, \bibinfo {author} {\bibfnamefont
  {D.}~\bibnamefont {Davidovikj}}, \bibinfo {author} {\bibfnamefont
  {H.}~\bibnamefont {van~der Zant}},\ and\ \bibinfo {author} {\bibfnamefont
  {P.}~\bibnamefont {Steeneken}},\ }\bibfield  {journal} {\bibinfo  {journal}
  {Microsyst. Nanoeng.}\ }\textbf {\bibinfo {volume} {6}},\ \href
  {https://doi.org/10.1038/s41378-020-00212-3} {10.1038/s41378-020-00212-3}
  (\bibinfo {year} {2020})\BibitemShut {NoStop}%
\bibitem [{\citenamefont {Fan}\ \emph {et~al.}(2019)\citenamefont {Fan},
  \citenamefont {Forsberg}, \citenamefont {Smith}, \citenamefont {Schröder},
  \citenamefont {Wagner}, \citenamefont {Östling}, \citenamefont {Lemme},\
  and\ \citenamefont {Niklaus}}]{Fan2019}%
  \BibitemOpen
  \bibfield  {author} {\bibinfo {author} {\bibfnamefont {X.}~\bibnamefont
  {Fan}}, \bibinfo {author} {\bibfnamefont {F.}~\bibnamefont {Forsberg}},
  \bibinfo {author} {\bibfnamefont {A.~D.}\ \bibnamefont {Smith}}, \bibinfo
  {author} {\bibfnamefont {S.}~\bibnamefont {Schröder}}, \bibinfo {author}
  {\bibfnamefont {S.}~\bibnamefont {Wagner}}, \bibinfo {author} {\bibfnamefont
  {M.}~\bibnamefont {Östling}}, \bibinfo {author} {\bibfnamefont {M.~C.}\
  \bibnamefont {Lemme}},\ and\ \bibinfo {author} {\bibfnamefont
  {F.}~\bibnamefont {Niklaus}},\ }\href
  {https://doi.org/10.1021/acs.nanolett.9b01759} {\bibfield  {journal}
  {\bibinfo  {journal} {Nano Letters}\ }\textbf {\bibinfo {volume} {19}},\
  \bibinfo {pages} {6788} (\bibinfo {year} {2019})},\ \bibinfo {note} {pMID:
  31478660},\ \Eprint
  {https://arxiv.org/abs/https://doi.org/10.1021/acs.nanolett.9b01759}
  {https://doi.org/10.1021/acs.nanolett.9b01759} \BibitemShut {NoStop}%
\bibitem [{\citenamefont {Zhou}\ \emph {et~al.}(2015)\citenamefont {Zhou},
  \citenamefont {Zheng}, \citenamefont {Onishi}, \citenamefont {Crommie},\ and\
  \citenamefont {Zettl}}]{Zhou2015}%
  \BibitemOpen
  \bibfield  {author} {\bibinfo {author} {\bibfnamefont {Q.}~\bibnamefont
  {Zhou}}, \bibinfo {author} {\bibfnamefont {J.}~\bibnamefont {Zheng}},
  \bibinfo {author} {\bibfnamefont {S.}~\bibnamefont {Onishi}}, \bibinfo
  {author} {\bibfnamefont {M.~F.}\ \bibnamefont {Crommie}},\ and\ \bibinfo
  {author} {\bibfnamefont {A.~K.}\ \bibnamefont {Zettl}},\ }\href
  {https://doi.org/10.1073/pnas.1505800112} {\bibfield  {journal} {\bibinfo
  {journal} {Proceedings of the National Academy of Sciences}\ }\textbf
  {\bibinfo {volume} {112}},\ \bibinfo {pages} {8942} (\bibinfo {year}
  {2015})},\ \Eprint
  {https://arxiv.org/abs/https://www.pnas.org/doi/pdf/10.1073/pnas.1505800112}
  {https://www.pnas.org/doi/pdf/10.1073/pnas.1505800112} \BibitemShut {NoStop}%
\bibitem [{\citenamefont {Wittmann}\ \emph {et~al.}(2019)\citenamefont
  {Wittmann}, \citenamefont {Glacer}, \citenamefont {Wagner}, \citenamefont
  {Pindl},\ and\ \citenamefont {Lemme}}]{Wittmann2019}%
  \BibitemOpen
  \bibfield  {author} {\bibinfo {author} {\bibfnamefont {S.}~\bibnamefont
  {Wittmann}}, \bibinfo {author} {\bibfnamefont {C.}~\bibnamefont {Glacer}},
  \bibinfo {author} {\bibfnamefont {S.}~\bibnamefont {Wagner}}, \bibinfo
  {author} {\bibfnamefont {S.}~\bibnamefont {Pindl}},\ and\ \bibinfo {author}
  {\bibfnamefont {M.~C.}\ \bibnamefont {Lemme}},\ }\href
  {https://doi.org/10.1021/acsanm.9b00998} {\bibfield  {journal} {\bibinfo
  {journal} {ACS Applied Nano Materials}\ }\textbf {\bibinfo {volume} {2}},\
  \bibinfo {pages} {5079} (\bibinfo {year} {2019})},\ \Eprint
  {https://arxiv.org/abs/https://doi.org/10.1021/acsanm.9b00998}
  {https://doi.org/10.1021/acsanm.9b00998} \BibitemShut {NoStop}%
\bibitem [{\citenamefont {Wood}\ \emph {et~al.}(2019)\citenamefont {Wood},
  \citenamefont {Torin}, \citenamefont {Al-mashaal}, \citenamefont {Smith},
  \citenamefont {Mastropaolo}, \citenamefont {Newton},\ and\ \citenamefont
  {Cheung}}]{Wood2019}%
  \BibitemOpen
  \bibfield  {author} {\bibinfo {author} {\bibfnamefont {G.~S.}\ \bibnamefont
  {Wood}}, \bibinfo {author} {\bibfnamefont {A.}~\bibnamefont {Torin}},
  \bibinfo {author} {\bibfnamefont {A.~K.}\ \bibnamefont {Al-mashaal}},
  \bibinfo {author} {\bibfnamefont {L.~S.}\ \bibnamefont {Smith}}, \bibinfo
  {author} {\bibfnamefont {E.}~\bibnamefont {Mastropaolo}}, \bibinfo {author}
  {\bibfnamefont {M.~J.}\ \bibnamefont {Newton}},\ and\ \bibinfo {author}
  {\bibfnamefont {R.}~\bibnamefont {Cheung}},\ }\href
  {https://doi.org/10.1109/JSEN.2019.2914401} {\bibfield  {journal} {\bibinfo
  {journal} {IEEE Sensors Journal}\ }\textbf {\bibinfo {volume} {19}},\
  \bibinfo {pages} {7234} (\bibinfo {year} {2019})}\BibitemShut {NoStop}%
\bibitem [{\citenamefont {Todorović}\ \emph {et~al.}(2015)\citenamefont
  {Todorović}, \citenamefont {Matković}, \citenamefont {Milićević},
  \citenamefont {Jovanović}, \citenamefont {Gajić}, \citenamefont {Salom},\
  and\ \citenamefont {Spasenović}}]{Todorovic2015}%
  \BibitemOpen
  \bibfield  {author} {\bibinfo {author} {\bibfnamefont {D.}~\bibnamefont
  {Todorović}}, \bibinfo {author} {\bibfnamefont {A.}~\bibnamefont
  {Matković}}, \bibinfo {author} {\bibfnamefont {M.}~\bibnamefont
  {Milićević}}, \bibinfo {author} {\bibfnamefont {D.}~\bibnamefont
  {Jovanović}}, \bibinfo {author} {\bibfnamefont {R.}~\bibnamefont {Gajić}},
  \bibinfo {author} {\bibfnamefont {I.}~\bibnamefont {Salom}},\ and\ \bibinfo
  {author} {\bibfnamefont {M.}~\bibnamefont {Spasenović}},\ }\href
  {https://doi.org/10.1088/2053-1583/2/4/045013} {\bibfield  {journal}
  {\bibinfo  {journal} {2D Materials}\ }\textbf {\bibinfo {volume} {2}},\
  \bibinfo {pages} {045013} (\bibinfo {year} {2015})}\BibitemShut {NoStop}%
\bibitem [{\citenamefont {Woo}\ \emph {et~al.}(2017)\citenamefont {Woo},
  \citenamefont {Han}, \citenamefont {Lee}, \citenamefont {Cho}, \citenamefont
  {Seong}, \citenamefont {Choi},\ and\ \citenamefont {Cho}}]{Woo2017}%
  \BibitemOpen
  \bibfield  {author} {\bibinfo {author} {\bibfnamefont {S.}~\bibnamefont
  {Woo}}, \bibinfo {author} {\bibfnamefont {J.-H.}\ \bibnamefont {Han}},
  \bibinfo {author} {\bibfnamefont {J.~H.}\ \bibnamefont {Lee}}, \bibinfo
  {author} {\bibfnamefont {S.}~\bibnamefont {Cho}}, \bibinfo {author}
  {\bibfnamefont {K.-W.}\ \bibnamefont {Seong}}, \bibinfo {author}
  {\bibfnamefont {M.}~\bibnamefont {Choi}},\ and\ \bibinfo {author}
  {\bibfnamefont {J.-H.}\ \bibnamefont {Cho}},\ }\href
  {https://doi.org/10.1021/acsami.6b12184} {\bibfield  {journal} {\bibinfo
  {journal} {ACS Applied Materials \& Interfaces}\ }\textbf {\bibinfo {volume}
  {9}},\ \bibinfo {pages} {1237} (\bibinfo {year} {2017})},\ \bibinfo {note}
  {pMID: 28055184},\ \Eprint
  {https://arxiv.org/abs/https://doi.org/10.1021/acsami.6b12184}
  {https://doi.org/10.1021/acsami.6b12184} \BibitemShut {NoStop}%
\bibitem [{\citenamefont {Li}\ \emph {et~al.}(2015)\citenamefont {Li},
  \citenamefont {Kinloch}, \citenamefont {Young}, \citenamefont {Novoselov},
  \citenamefont {Anagnostopoulos}, \citenamefont {Parthenios}, \citenamefont
  {Galiotis}, \citenamefont {Papagelis}, \citenamefont {Lu},\ and\
  \citenamefont {Britnell}}]{Li2015}%
  \BibitemOpen
  \bibfield  {author} {\bibinfo {author} {\bibfnamefont {Z.}~\bibnamefont
  {Li}}, \bibinfo {author} {\bibfnamefont {I.~A.}\ \bibnamefont {Kinloch}},
  \bibinfo {author} {\bibfnamefont {R.~J.}\ \bibnamefont {Young}}, \bibinfo
  {author} {\bibfnamefont {K.~S.}\ \bibnamefont {Novoselov}}, \bibinfo {author}
  {\bibfnamefont {G.}~\bibnamefont {Anagnostopoulos}}, \bibinfo {author}
  {\bibfnamefont {J.}~\bibnamefont {Parthenios}}, \bibinfo {author}
  {\bibfnamefont {C.}~\bibnamefont {Galiotis}}, \bibinfo {author}
  {\bibfnamefont {K.}~\bibnamefont {Papagelis}}, \bibinfo {author}
  {\bibfnamefont {C.-Y.}\ \bibnamefont {Lu}},\ and\ \bibinfo {author}
  {\bibfnamefont {L.}~\bibnamefont {Britnell}},\ }\href
  {https://doi.org/10.1021/nn507202c} {\bibfield  {journal} {\bibinfo
  {journal} {ACS Nano}\ }\textbf {\bibinfo {volume} {9}},\ \bibinfo {pages}
  {3917} (\bibinfo {year} {2015})},\ \bibinfo {note} {pMID: 25765609},\ \Eprint
  {https://arxiv.org/abs/https://doi.org/10.1021/nn507202c}
  {https://doi.org/10.1021/nn507202c} \BibitemShut {NoStop}%
\bibitem [{\citenamefont {Baglioni}\ \emph {et~al.}(2023)\citenamefont
  {Baglioni}, \citenamefont {Pezone}, \citenamefont {Vollebregt}, \citenamefont
  {Cvetanović~Zobenica}, \citenamefont {Spasenović}, \citenamefont
  {Todorović}, \citenamefont {Liu}, \citenamefont {Verbiest}, \citenamefont
  {van~der Zant},\ and\ \citenamefont {Steeneken}}]{Baglioni2023}%
  \BibitemOpen
  \bibfield  {author} {\bibinfo {author} {\bibfnamefont {G.}~\bibnamefont
  {Baglioni}}, \bibinfo {author} {\bibfnamefont {R.}~\bibnamefont {Pezone}},
  \bibinfo {author} {\bibfnamefont {S.}~\bibnamefont {Vollebregt}}, \bibinfo
  {author} {\bibfnamefont {K.}~\bibnamefont {Cvetanović~Zobenica}}, \bibinfo
  {author} {\bibfnamefont {M.}~\bibnamefont {Spasenović}}, \bibinfo {author}
  {\bibfnamefont {D.}~\bibnamefont {Todorović}}, \bibinfo {author}
  {\bibfnamefont {H.}~\bibnamefont {Liu}}, \bibinfo {author} {\bibfnamefont
  {G.~J.}\ \bibnamefont {Verbiest}}, \bibinfo {author} {\bibfnamefont
  {H.~S.~J.}\ \bibnamefont {van~der Zant}},\ and\ \bibinfo {author}
  {\bibfnamefont {P.~G.}\ \bibnamefont {Steeneken}},\ }\href
  {https://doi.org/10.1039/D2NR05147H} {\bibfield  {journal} {\bibinfo
  {journal} {Nanoscale}\ }\textbf {\bibinfo {volume} {15}},\ \bibinfo {pages}
  {6343} (\bibinfo {year} {2023})}\BibitemShut {NoStop}%
\bibitem [{\citenamefont {Wang}\ \emph {et~al.}(2015)\citenamefont {Wang},
  \citenamefont {Zou}, \citenamefont {Song},\ and\ \citenamefont
  {Tao}}]{Wang2015}%
  \BibitemOpen
  \bibfield  {author} {\bibinfo {author} {\bibfnamefont {Z.}~\bibnamefont
  {Wang}}, \bibinfo {author} {\bibfnamefont {Q.}~\bibnamefont {Zou}}, \bibinfo
  {author} {\bibfnamefont {Q.}~\bibnamefont {Song}},\ and\ \bibinfo {author}
  {\bibfnamefont {J.}~\bibnamefont {Tao}},\ }\href@noop {} {\bibfield
  {journal} {\bibinfo  {journal} {2015 Transducers-2015 18th International
  Conference on Solid-State Sensors, Actuators and Microsystems (IEEE
  TRANSDUCERS)}\ ,\ \bibinfo {pages} {375}} (\bibinfo {year}
  {2015})}\BibitemShut {NoStop}%
\bibitem [{\citenamefont {Liu}\ \emph {et~al.}(2023{\natexlab{b}})\citenamefont
  {Liu}, \citenamefont {Basuvalingam}, \citenamefont {Lodha}, \citenamefont
  {Bol}, \citenamefont {van~der Zant}, \citenamefont {Steeneken},\ and\
  \citenamefont {Verbiest}}]{Liu2023_ALD}%
  \BibitemOpen
  \bibfield  {author} {\bibinfo {author} {\bibfnamefont {H.}~\bibnamefont
  {Liu}}, \bibinfo {author} {\bibfnamefont {S.~B.}\ \bibnamefont
  {Basuvalingam}}, \bibinfo {author} {\bibfnamefont {S.}~\bibnamefont {Lodha}},
  \bibinfo {author} {\bibfnamefont {A.~A.}\ \bibnamefont {Bol}}, \bibinfo
  {author} {\bibfnamefont {H.~S.~J.}\ \bibnamefont {van~der Zant}}, \bibinfo
  {author} {\bibfnamefont {P.~G.}\ \bibnamefont {Steeneken}},\ and\ \bibinfo
  {author} {\bibfnamefont {G.~J.}\ \bibnamefont {Verbiest}},\ }\href
  {https://doi.org/10.1088/2053-1583/acf58a} {\bibfield  {journal} {\bibinfo
  {journal} {2D Materials}\ }\textbf {\bibinfo {volume} {10}},\ \bibinfo
  {pages} {045023} (\bibinfo {year} {2023}{\natexlab{b}})}\BibitemShut
  {NoStop}%
\bibitem [{\citenamefont {Orlando}(2008)}]{Orlando2008}%
  \BibitemOpen
  \bibfield  {author} {\bibinfo {author} {\bibfnamefont {P.~B. . C.~G.}\
  \bibnamefont {Orlando}, \bibfnamefont {S.}},\ }\href@noop {} {\bibfield
  {journal} {\bibinfo  {journal} {Proceedings of the ISMA 2008 International
  Conference on Noise and Vibration Engineering}\ ,\ \bibinfo {pages} {229}}
  (\bibinfo {year} {2008})}\BibitemShut {NoStop}%
\bibitem [{\citenamefont {Fletcher}\ and\ \citenamefont
  {Munson}(1933)}]{Fletcher1933}%
  \BibitemOpen
  \bibfield  {author} {\bibinfo {author} {\bibfnamefont {H.}~\bibnamefont
  {Fletcher}}\ and\ \bibinfo {author} {\bibfnamefont {W.}~\bibnamefont
  {Munson}},\ }\href@noop {} {\bibfield  {journal} {\bibinfo  {journal} {J.
  Acoust. Soc. Am.}\ ,\ \bibinfo {pages} {82}} (\bibinfo {year}
  {1933})}\BibitemShut {NoStop}%
\bibitem [{\citenamefont {for Sound Level Meters Z24.3-1936 for Measurement~of
  Noise}\ and\ \citenamefont {Sounds}(1936)}]{ATS1936}%
  \BibitemOpen
  \bibfield  {author} {\bibinfo {author} {\bibfnamefont {A.~T.~S.}\
  \bibnamefont {for Sound Level Meters Z24.3-1936 for Measurement~of Noise}}\
  and\ \bibinfo {author} {\bibfnamefont {O.}~\bibnamefont {Sounds}},\
  }\href@noop {} {\bibfield  {journal} {\bibinfo  {journal} {J. Acoust. Soc.
  Am.}\ ,\ \bibinfo {pages} {147}} (\bibinfo {year} {1936})}\BibitemShut
  {NoStop}%
\bibitem [{\citenamefont
  {https://nl.mathworks.com/matlabcentral/fileexchange/69
  octave}()}]{Octave69}%
  \BibitemOpen
  \bibfield  {author} {\bibinfo {author} {\bibnamefont
  {https://nl.mathworks.com/matlabcentral/fileexchange/69 octave}},\
  }\href@noop {} {}\BibitemShut {NoStop}%
\end{thebibliography}%

\onecolumngrid

\newpage

\begin{center}
    
\section*{The graphene squeeze-film microphone - Supplementary Information}

M.P.~Abrahams$^\dagger$, J.~Martinez$^\ddagger$, P.G.~Steeneken$^\dagger$, G.J.~Verbiest$^{*\dagger}$\\ \ \\

$^\dagger$ {\it Department of Precision and Microsystems Engineering, Delft University of Technology, Mekelweg 2, 2628 CD Delft, The Netherlands, EU}\\
$^\ddagger$ {\it Multimedia Computing Group, Intelligent Systems Department, Faculty of Electrical Engineering, Mathematics and Computer Science, Delft University of Technology, 2628 XE Delft, The Netherlands, EU}\\ \ \\

E-mail: G.J.Verbiest@tudelft.nl
\end{center}

\setcounter{equation}{0}
\renewcommand{\theequation}{S\arabic{equation}}
\setcounter{figure}{0}
\renewcommand{\thefigure}{S\arabic{figure}}

\subsection*{1. Fabrication and characterization\label{subs0}}

\begin{figure}[b]
\begin{center}
\includegraphics[draft=false,keepaspectratio=true,clip,width=1.0\linewidth]{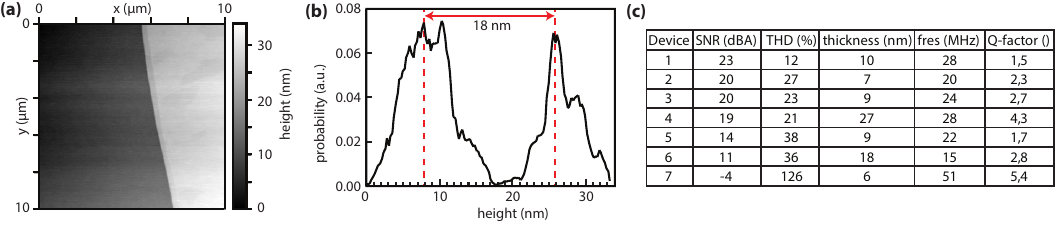} 
\end{center}
\caption{(a) Atomic force microscopy (AFM) topography image of graphene membrane on device 6. (b) normalized height distribution of the AFM image shown in panel (a). The dashed red lines indicate the center of the peaks corresponding to the surface of the substrate and of the membrane. The distance of 18 nm between the dashed red lines corresponds to the thickness of the membrane. (c) table of the determined signal-to-noise ratio (SNR), total harmonic distortion (THD), thickness (nm), resonance frequency (MHz), and Q-factor of the measured devices. 
\label{figS0}}
\end{figure}

\begin{figure}[t]
\begin{center}
\includegraphics[draft=false,keepaspectratio=true,clip,width=0.5\linewidth]{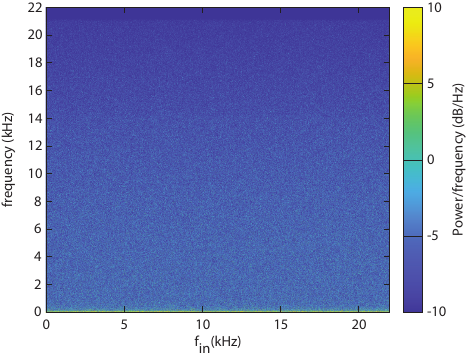} 
\end{center}
\caption{Spectrogram recorded with the graphene squeeze-film microphone as described in the main text when glued upon and directly excited with a piezo-electric transducer. The absence of a signal indicates the graphene squeeze-film microphone is insensitive to sound and vibrations coupling in trhrough the substrate.
\label{figS1}}
\end{figure}

A Si wafer with 285~\si{nm} dry SiO$_{\rm 2}$ is spin coated with positive e-beam resist and exposed by electron-beam lithography. Afterwards, the SiO$_{\rm 2}$ layer without protection is completely etched using CHF$_{\rm 3}$ and Ar plasma in a reactive ion etcher. The edges of dumbbells are examined to be well-defined by scanning electron microscopy (SEM) and AFM. After resist removal, 2D nanoflakes are exfoliated by Scotch tape, and then separately transferred onto the substrate at room temperature through a deterministic dry stamping technique, as detailed in our earlier work \cite{Liu2023_ALD}. Using tapping mode atomic force microscopy (AFM), we measure the height difference between the membrane and the Si/SiO$_{\rm 2}$ substrate. As Fig.~\ref{figS0} shows, we find a membrane thickness $t$ of 18~\si{nm} for device 6.

\subsection*{2. Transfer function reference microphone and speaker\label{subs1}}

The response of the graphene membranes and the reference microphone (AKG C417-L), for example in Fig.~\ref{fig2}c of the main text, are convoluted with the transfer function of the speaker (Jabra speak 510).
Neither the transfer function of the reference microphone and the speaker are not in detail available, as they also depend on the exact placement with respect to each other.
Therefore, we used a well-calibrated Genelec 8020 GPM speaker to estimate the transfer functions.
We performed measurements with the Genelec 8020 GPM speaker placed on its tripod and the Jabra speak 510 suspended in mid-air using thin wires.
The microphone is also suspended in mid-air using thin wires and is placed at the same distances from the speaker as during the graphene measurements.
All devices are attached to a FireFace UFX+ soundcard for the read-out and actuation.
We use a frequency sine sweep method \cite{Orlando2008} to actuate the speakers and measure with the microphone.
As the transfer function of the Genelec 8020 GPM speaker is known, we determine the transfer function of the AKG C417-L.
In turn, this allows us to determine the transfer function of the Jabra speak 510.
This information is used to deconvolute the speaker response from the graphene response in Fig.~\ref{fig2}c and \ref{fig2}d.


\subsection*{3. Experimental signal-to-noise ratio and total harmonic distortion\label{subs2}}

We determined the signal-to-noise ratio and total harmonic distortion for all graphene devices. 
Following state-of-the-art procedures for MEMS microphones, we specify the signal-to-noise ratio for an acoustic excitation of 80~dB$_\textrm{SPL}$ at 1~kHz.
First, we make a spectrogram of the data recorded from the reference microphone and the graphene device while sweeping $f_{\rm in}$ from 20~Hz to 22~kHz.
Using the spectrogram of the reference microphone, we create a mask for the signal at $f_{\rm in}$ at 1~kHz and at $n$ integer multiples of $f_{\rm in}$.
The mask at $f_{\rm in}$ is used to identify the signal in the spectrogram of the graphene device and the ones at $n f_{\rm in}$ to determine the total harmonic distortion.
We then select a small band around $f_{\rm in} = 1$~kHz of 350~Hz that is not affected by the phase lock loss of the PLL and is thus representative for the signal.
We repeat this procedure for small bands around $n$ integer multiples of $f_{\rm in}$, which is thus representative for the total harmonic distortion.
The remaining parts of the spectrogram of the graphene device represent the left-over noise.
Then, we make an inverse Fourier transform of the identified signal, total harmonic distortion, and the left-over noise.
Utilising the time domain results we acquire the signal, total harmonic distortion, and noise powers in dBA, in accordance with the theory as stated in \cite{Fletcher1933,ATS1936} and using the Matlab code which can be found at \cite{Octave69}. 
The resulting signal, total harmonic distortion, and noise powers and then averaged to find a global value for the entire measurement.
Finally, we subtract the signal and noise powers to acquire the signal-to-noise ratio.
We find the total harmonic distortion by the total harmonic distortion power by the sum of the total harmonic distortion power and the signal power and then convert this number from dBA into a percentage.

\subsection*{4. Thermo-mechanical limit of the signal-to-noise ratio\label{subs3}}

The squeeze-film effect relates the resonance frequency $f_r$ of the graphene membrane to the ambient pressure $P_{\rm amb}$ and the distance $g_{\rm 0}$ to the back-plate:

\begin{equation}%
f_{\rm res}^2=f_{\rm 0}^2 + \frac{P_{\rm amb}}{4\pi^2 g_{\rm 0} \rho h}.%
\label{seq1}%
\end{equation}%

Here, $\rho$ is the mass density of the graphene and $h$ its thickness.
It is important to realize that the motion $\delta x(t)$ as a function of time $t$ of the graphene membrane itself also modulates the distance $g_{\rm 0}$.
The true distance between the graphene and the back-plate is given by $g_{\rm 0}+\delta x(t)$.
According to equation~\ref{seq1}, this will result in frequency fluctuation and thereby set a limit on the performance of squeeze-film microphones. 
We can estimate this limit by making a Taylor expansion of equation~\ref{seq1} in $\delta x(t)$:

\begin{equation}%
f_{\rm res}^2 + 2f_{\rm res}\delta f(t) + \delta f^2 = f_{\rm 0}^2 + 2f_{\rm 0}\delta f_0(t) + \delta f_{\rm 0}^2(t) + \frac{P_{\rm amb}}{4\pi^2 g_{\rm 0} \rho h} \left(1 - \frac{\delta x(t)}{g_{\rm 0}} + \frac{\delta x(t)^2}{g_{\rm 0}^2} \right).%
\label{seq2}%
\end{equation}%
In this expansion, we already included the frequency fluctuations $\delta f_{\rm res}(t)$ and $\delta f_{\rm 0}(t)$. The former represents the frequency fluctuations of the squeeze-film microphone and the latter the frequency fluctuations in vacuum conditions.

In order to quantify the resulting frequency fluctuations $\delta f_{\rm res}(t)$, we take the time average of equation~\ref{seq2} and subtract equation~\ref{seq1} to find:

\begin{equation}%
<\delta f^2> = <\delta f_{\rm 0}^2(t)> + \frac{P_{\rm amb}}{4\pi^2 g_{\rm 0} \rho h} \frac{<\delta x(t)^2>}{g_{\rm 0}^2}.%
\label{seq3}%
\end{equation}%
Here, $<...>$ denotes a time-averaged value. In this step, we assumed that $<x(t)>$, $<\delta f_{\rm res}(t)>$ and $<\delta f_{\rm 0}(t)>$ are zero.

From equipartition theorem, we find an analytical expression for $<\delta x(t)^2>$:

\begin{equation}%
<\delta x(t)^2> = \frac{k_{\rm B} T}{4 \pi^2 m_{\rm eff} f_{\rm res}^2},%
\label{seq4}%
\end{equation}%
where the mass $m_{\rm eff}$ is the effective mass of the fundamental mode of the membrane and is given by $m_{\rm eff} = \gamma \rho h \pi r^2$, in which $\gamma = 0.269$. We denote the radius of the membrane with $r$.

By inserting equation~\ref{seq4} in equation~\ref{seq3} and setting $<\delta f_{\rm 0}^2(t)>$ equal to zero, we find a fundamental limit for the frequency fluctuations $<\delta f^2>$ of a squeeze-film microphone:

\begin{equation}%
<\delta f^2> = \frac{P_{\rm amb} k_{\rm B} T \gamma \pi r^2}{16\pi^4 g_{\rm 0}^3 m_{\rm eff}^2 f_{\rm res}^2}.%
\label{seq5}%
\end{equation}%

In order to estimate the ultimate signal-to-noise ratio of squeeze-film microphone, we also need the sensitivity $S$:

\begin{equation}%
S = \frac{\partial f_{\rm res}}{\partial P_{\rm amb}} = \frac{1}{8\pi^2 g_{\rm 0} \rho h f_{\rm res}} = \frac{\gamma \pi r^2 f_{\rm res}}{2 k g_{\rm 0}}.%
\label{seq6}%
\end{equation}%

The second equality in Eq.~\ref{seq6} arises from $2\pi f_{\rm res} = \sqrt{k/m_{\rm eff}}$, in which $1/k$ is the compliance of the membrane. For commercial microphones, the signal-to-noise ratio is commonly specified at a sound pressure level of 80 dB$_\textrm{SPL}$, which corresponds to a pressure wave with an amplitude of 0.2 Pa. The signal one expects from the squeeze-film microphone is then $0.2 S$. When denoting the noise source by $N$, we find a SNR ratio of:

\begin{equation}%
SNR_{\rm limit} = 20\log\left(\frac{0.2 S}{N}\right) = 20\log\left(\frac{0.2 \gamma \pi r^2 f_{\rm res}}{2 k g_{\rm 0} N}\right).%
\label{seq7}%
\end{equation}%

\noindent
In case $N$ is set by the thermo-mechanical noise, we find the ultimate noise limit is given by $\sqrt{<\delta f^2>}$. This leads to the following signal-to-noise ratio in dB:

\begin{equation}%
SNR_{\rm limit} = 20\log\left(\frac{0.2 S}{\sqrt{<\delta f^2>}}\right) = 20\log\left(\frac{0.1 \sqrt{\gamma \pi r^2 g_{\rm 0}}}{\sqrt{P_{\rm amb} k_{\rm B} T}}\right).%
\label{seq8}%
\end{equation}%

\noindent
The ultimate signal-to-noise ratio solely depends on the radius of the membrane, the ambient pressure, and the temperature. 

\end{document}